# Data Mining: A prediction for performance improvement using classification

Brijesh Kumar Bhardwaj
Research Scholar, Singhaniya University,
Rajasthan, India
wwwbkb@rediffmail.com

Saurabh Pal
Dept. of Computer Applications,
VBS Purvanchal University,
Jaunpur (UP) - 224001, India
drsaurabhpal@yahoo.co.in

*Abstract*—Now-a-days the amount of data stored in educational database increasing rapidly. These databases contain hidden information for improvement of students' performance. The performance in higher education in India is a turning point in the academics for all students. This academic performance is influenced by many factors, therefore it is essential to develop predictive data mining model for students' performance so as to identify the difference between high learners and slow learners student.

In the present investigation, an experimental methodology was adopted to generate a database. The raw data was preprocessed in terms of filling up missing values, transforming values in one form into another and relevant attribute/ variable selection. As a result, we had 300 student records, which were used for by Byes classification prediction model construction.

## I. INTRODUCTION

The ability to predict a student's performance is very important in educational environments. Students' academic performance is based upon diverse factors like personal, social, psychological and other environmental variables. A very promising tool to attain this objective is the use of Data Mining. Data mining techniques are used to operate on large amount of data to discover hidden patterns and relationships helpful in decision making.

In fact, one of the most useful data mining techniques in e-learning is classification. Classification is a predictive data mining technique, makes prediction about values of data using known results found from different data [1]. Predictive models have the specific aim of allowing us to predict the unknown values of variables of interest given known values of other variables. Predictive modeling can be thought of as learning a mapping from an input set of vector measurements to a scalar output [4]. Classification maps data into predefined groups of classes. It is often referred to as supervised learning because the classes are determined before examining the data.

Prediction models that include all personal, social, psychological and other environmental variables are necessitated for the effective prediction of the performance of the students. The prediction of student performance with high accuracy is beneficial for identify the students with low academic achievements initially. It is required that the identified students can be assisted more by the teacher so that their performance is improved in future.

In this connection, the objectives of the present investigation were framed so as to assist the low academic achievers in higher education and they are:

(a) Generation of a data source of predictive variables,

(b) Identification of different factors, which effects a student's learning behavior and performance during academic career

(c) Construction of a prediction model using classification data mining techniques on the basis of identified predictive variables and

(d) Validation of the developed model for higher education students studying in Indian Universities or Institutions.

## II. BACKGROUND AND RELATED WORK

Data Mining can be used in educational field to enhance our understanding of learning process to focus on identifying, extracting and evaluating variables related to the learning process of students as described by Alaa el-Halees [2]. Mining in educational environment is called Educational Data Mining.

Han and Kamber [6] describes data mining software that allow the users to analyze data from different dimensions, categorize it and summarize the relationships which are identified during the mining process.

Pandey and Pal [10] conducted study on the student performance based by selecting 600 students from different colleges of Dr. R. M. L. Awadh University, Faizabad, India. By means of Bayes Classification on category, language and background qualification, it was found that whether new comer students will performer or not.

Hijazi and Naqvi [7] conducted as study on the student performance by selecting a sample of 300 students (225 males, 75 females) from a group of colleges affiliated to Punjab university of Pakistan. The hypothesis that was stated as "Student's attitude towards attendance in class, hours spent in study on daily basis after college, students' family income, students' mother's age and mother's education are significantly related with student performance" was framed. By means of simple linear regression analysis, it was found that the factors like mother's education and student's family income were highly correlated with the student academic performance.

Khan [8] conducted a performance study on 400 students comprising 200 boys and 200 girls selected from the senior secondary school of Aligarh Muslim University, Aligarh, India with a main objective to establish the prognostic value of different measures of cognition, personality and demographic variables for success at higher secondary level in science stream. The selection was based on cluster sampling technique in which the entire population of interest was divided into groups, or clusters, and a random sample of these clusters was selected for further analyses. It was found that girls with high socio-economic status had relatively higher academic achievement in science stream and boys with low socio-economic status had relatively higher academic achievement in general.

Galit [5] gave a case study that use students data to analyze their learning behavior to predict the results and to warn students at risk before their final exams.

Al-Radaideh, et al [1] applied a decision tree model to predict the final grade of students who studied the C++ course in Yarmouk University, Jordan in the year 2005. Three different classification methods namely ID3, C4.5, and the NaïveBayes were used. The outcome of their results indicated that Decision Tree model had better prediction than other models.

Pandey and Pal [11] conducted study on the student performance based by selecting 60 students from a degree college of Dr. R. M. L. Awadh University, Faizabad, India. By means of association rule they find the interestingness of student in opting class teaching language.

Bray [2], in his study on private tutoring and its implications, observed that the percentage of students receiving private tutoring in India was relatively higher than in Malaysia, Singapore, Japan, China and Sri Lanka. It was also observed that there was an enhancement of academic performance with the intensity of private tutoring and this variation of intensity of private tutoring depends on the collective factor namely socio-economic conditions.

### III. DATA MINING PROCESS

In this study, data gathered from different degree colleges and institutions affiliated with Dr. R. M. L. Awadh University, Faizabad, India. These data are analyzed using classification method to predict the student's performance. In order to apply this technique following steps are performed in sequence:

#### A. Data Preparations

The data set used in this study was obtained from different colleges on the sampling method of computer Applications department of course BCA (Bachelor of Computer Applications) of session 2009-10. Initially size of the data is 300. In this step data stored in different tables was joined in a single table after joining process errors were removed.

#### B. Data selection and transformation

In this step only those fields were selected which were required for data mining. A few derived variables were selected. While some of the information for the variables was extracted from the database. All the predictor and response variables which were derived from the database are given in Table 1 for reference.

Table 1: Student Related Variables

| Variable | Description | Possible Values |
|---|---|---|
| Sex | Students Sex | {Male, Female} |
| Cat | Students category | {General, OBC, SC, ST} |
| Med | Medium of Teaching | {Hindi, English, Mix} |
| SFH | Students food habit | {veg, non-veg} |
| SOH | Students other habit | {drinking, smoking, both, not-applicable} |
| LLoc | Living Location | {Village, Town, Tahseel, District} |
| Hos | Student live in hostel or not | {Yes, No} |
| FSize | student's family size | {1, 2, 3, >3} |
| FStat | Students family status | {Joint, Individual} |
| FAIn | Family annual income status | {BPL, poor, medium, high} |
| GSS | Students grade in Senior Secondary education | {O – 90% -100%, A – 80% - 89%, B – 70% - 79%, C – 60% - 69%, D – 50% - 59%, E – 40% - 49%, F - < 40%} |
| TColl | Students College Type | {Female, Co-education} |
| FQual | Fathers qualification | {no-education, elementary, secondary, graduate, post-graduate, doctorate, not-applicable} |
| MQual | Mother's Qualification | {no-education, |
| | | elementary, secondary, graduate, post-graduate, doctorate, not-applicable} |
| FOcc | Father's Occupation | {Service, retired, not-applicable} |
| MOcc | Mother's Occupation | {House-wife, Service, retired, not-applicable} |
| GObt | Grade obtained in BCA | {First > 60% Second >45 & <60% Third >36 & <45% Fail < 36%} |

The domain values for some of the variables were defined for the present investigation as follows:

- **Cat –** From ancient time Indians are divided in many categories. These factors play a direct and indirect role in the daily lives including the education of young people. Admission process in India also includes different percentage of seats reserved for different categories. In terms of social status, the Indian population is grouped into four categories: General, Other Backward Class (OBC), Scheduled Castes (SC) and Scheduled Tribes (ST). Possible values are *General, OBC, SC and ST.*

- **Med –** This paper study covers only the degree colleges and institutions of Uttar Pradesh state of India. Here, medium of instructions are *Hindi* or *English* or *Mix* (Both Hindi and English).

- **SOH –** In modern society bad habits are increasing fast among college students. Here students other habit include *Drinking, Smoking, Both* or *Not-applicable*.

- **FSize-**. According to population statistics of India, the average number of children in a family is 3.1. Therefore, the maximum family size is fixed as 10 and possible range of values is from *one* to *ten*.

- **GSS -** Students grade in Senior Secondary education. Students who are in state board appear for five subjects each carry 100 marks. Grade are assigned to all students using following mapping *O – 90% to 100%, A – 80% - 89%, B – 70% - 79%, C – 60% - 69%, D – 50% - 59%, E – 40% - 49%,* and *F - < 40%}.*

- **GObt -** Marks/Grade obtained in BCA course and it is declared as response variable. It is also split into five class values: First – >60% , Second – >45% and <60%, Third – >36% and < 45%, Fail < 40%.

### C. Implementation of Mining Model

Various algorithms and techniques like Classification, Clustering, Regression, Artificial Intelligence, Neural Networks, Association Rules, Decision Trees, Genetic Algorithm, Nearest Neighbor method etc., are used for knowledge discovery from databases.

Classification is one of the most frequently studied problems by data mining and machine learning (ML) researchers. It consists of predicting the value of a (categorical) attribute (the class) based on the values of other attributes (the predicting attributes). There are different classification methods. In the present study we use the Bayesian Classification algorithm.

Bayes classification has been proposed that is based on Bayes rule of conditional probability. Bayes rule is a technique to estimate the likelihood of a property given the set of data as evidence or input Bayes rule or Bayes theorem is-

$$P(h_i | x_i) = \frac{P(x_i | h_i)P(h_i)}{P(x_i | h_i) + P(x_i | h_2)P(h_2)}$$

The approach is called "naïve" because it assumes the independence between the various attribute values. Naïve Bayes classification can be viewed as both a descriptive and a predictive type of algorithm. The probabilities are descriptive and are then used to predict the class membership for a target tuple. The naïve Bayes approach has several advantages: it is easy to use; unlike other classification approaches only one scan of the training data is required; easily handle mining value by simply omitting that probability [11]. An advantage of the naive Bayes classifier is that it requires a small amount of training data to estimate the parameters (means and variances of the variables) necessary for classification. Because independent variables are assumed, only the variances of the variables for each class need to be determined and not the entire covariance matrix. In spite of their naive design and apparently over-simplified assumptions, naive Bayes classifiers have worked quite well in many complex real-world situations.

For the present study, we selected five degree colleges running BCA course affiliated with Dr. R. M. L. Awadh University, Faizabad, UP, India. Out of five degree colleges two was an urban-based, unaided and co-educational school, the other one was a rural-based, aided and female college and the other two was rural-based, aided and co-education college. A total of 300 (226 males, 74 females) students of BCA course from these five colleges who appeared in 2010 examination were the samples for our study. All the information related to student's demographic, academic and socio-economic variables was obtained from the 300 students directly through questionnaire and University database. The mark obtained of these students was collected from the University Examination cell.

Given a training set the naïve Bayes algorithm first estimates the prior probability $P(C_j)$ for each class by counting how often each class occurs in the training data. For ach attribute value $x_i$ can be counted to determine $P(x_i)$. Similarly the probability $P(x_i | C_j)$ can be estimated by counting how often each value occurs in the class in the training data.

When classifying a target tuple, the conditional and prior probabilities generated from the training set are used to make the prediction. Then estimate $P(t_i | C_j)$ by

$$P(t_i | c_j) = \prod_{k=1}^{p}(x_{ij} | c_j)$$

To calculate $P(t_i)$ we can estimate the likelihood that $t_i$ is in each class. The probability that $t_i$ is in a class is the product of the conditional probabilities for each attribute value. The class with the highest probability is the one chosen for the tuple [10].

The present investigation used data mining as a tool with naïve Bayes classification algorithm as a technique to design the student performance prediction model. Filtered feature selection technique was used to select the best subset of variables on the basis of the values of probabilities.

### D. Result and Discussion

In the present study, those variables whose probability values were greater than 0.50 were given due considerations and the highly influencing variables with high probability values have been shown in Table 2. These features were used for prediction model construction. For both variable selection and prediction model construction, we have used MatLab.

Table 2: High Potential Variables

| Variable | Description | Probability |
|---|---|---|
| GSS | Students grade in Senior Secondary education | .8642 |
| LLoc | Living Location | .7862 |
| Med | Medium of Teaching | .7225 |
| MQual | Mother's Qualification | .6788 |
| SOH | Students other habit | .6653 |
| FAIn | Family annual income status | .5672 |
| FStat | Students family status | .5225 |

From the table 2, it is found that the students' performance is highly dependent on their grade obtained in Senior Secondary Examination, which is shown in Fig 1.

Figure 1: Relationship between GSS and GObt

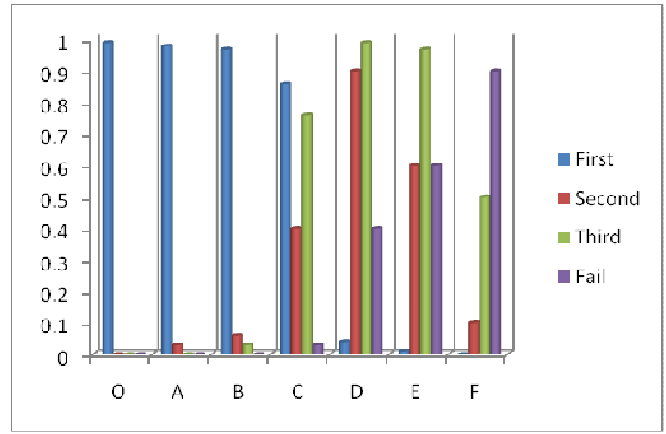

From the table 2, it is found that the second high potential variable for students' performance is their living location. The relationship between students living area and their grade obtained in BCA examination is shown in Fig 2.

Figure 2: Relationship between LLoc and GObt

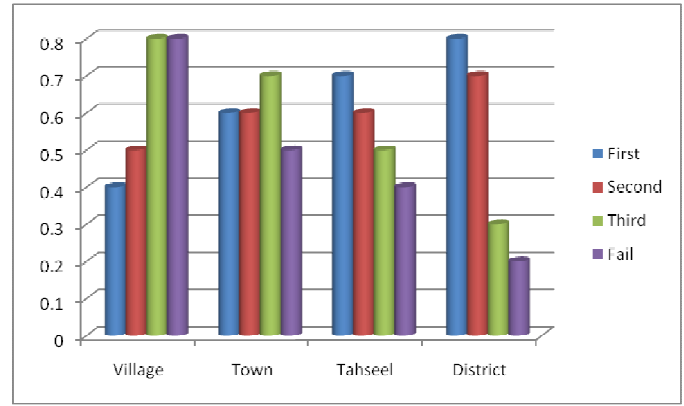

From the table 2, it is found that the third high potential variable for students' performance is medium of teaching. In Uttar Pradesh the mother tong language of students if Hindi. In Mixed and Hindi language students are more comfortable than English language. The relationship between students' medium of teaching and their grade obtained in BCA examination is shown in Fig 3.

Similarly, from table 2, it is found that Mother's Qualification, Students Other Habit, Family annual income and students' family status are other high potential variables that effect students' performance for obtaining higher grade in final examination.

Figure 3: Relationship between LLoc and GObt

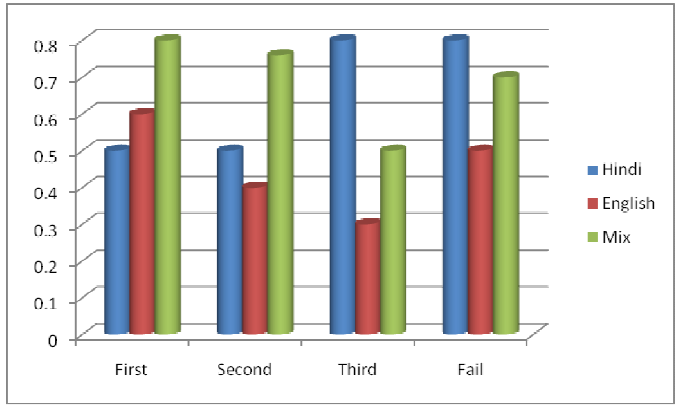

## IV Conclusion

In this paper, Bayesian classification method is used on student database to predict the students division on the basis of previous year database. This study will help to the students and the teachers to improve the division of the student. This study will also work to identify those students which needed special attention to reduce failing ration and taking appropriate action at right time.

Present study shows that academic performances of the students are not always depending on their own effort. Our investigation shows that other factors have got significant influence over students' performance. This proposal will improve the insights over existing methods.

## REFERENCES

[1]  AI-Radaideh,Q. A., AI-Shawakfa, E.M., and AI-Najjar, M. I., "Mining Student Data using Decision Trees", International Arab Conference on Information Technology(ACIT'2006), Yarmouk University, Jordan, 2006.
[2]  Alaa el-Halees, "Mining Students Data to Analyze e-Learning Behavior: A Case Study", 2009.
[3]  Bray, M. The Shadow Education System: Private Tutoring And Its Implications For Planners, (2nd ed.), UNESCO, PARIS, France, 2007.
[4]  David Hand, Heikki, Mannil Padraic smyth, "Principles of Data Mining" PHI
[5]  Galit.et.al, "Examining online learning processes based on log files analysis: a case study".  Research, Reflection and Innovations in Integrating ICT in Education 2007.
[6]  Han,J. and Kamber, M., "Data Mining: Concepts and Techniques", 2nd edition. The Morgan Kaufmann Series in Data Management Systems, Jim Gray, Series Editor, 2006.
[7]  Hijazi, S. T., and Naqvi, R.S.M.M., "Factors Affecting Student's Performance: A Case of Private Colleges", Bangladesh e-Journal of Sociology, Vol. 3, No. 1, 2006.
[8]  Khan, Z. N., "Scholastic Achievement of Higher Secondary Students in Science Stream", Journal of Social Sciences, Vol. 1, No. 2, 2005, pp. 84-87.
[9]  Margret H. Dunham, "Data Mining: Introductory and advance topic".
[10] Pandey, U. K. and Pal, S., "Data Mining: A prediction of performer or underperformer using classification", (IJCSIT) International Journal of Computer Science and Information Technology, Vol. 2(2), 2011, 686-690, ISSN:0975-9646.
[11] Pandey, U. K. and Pal, S., "A Data Mining View on Class Room Teaching Language", (IJCSI) International Journal of Computer Science Issue, Vol. 8, Issue 2, March -2011, 277-282, ISSN:1694-0814
[12] Westphal, C., Blaxton, T., "Data Mining Solutions", John Wiley, 2008.

**Author Profile**

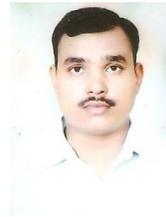

**Brijesh Kumar Bhardwaj** is Assistant Professor in the Department of  Computer Applications, Dr. R. M. L. Avadh University Faizabad India. He obtained his M.C.A degree from Dr. R. M. L. Avadh University Faizabad (2003) and M.Phil. in Computer Applications from Vinayaka mission University, Tamilnadu. He is currently doing research in Data Mining and Knowledge Discovery.

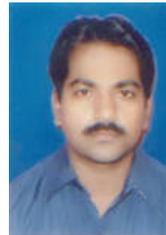

**Saurabh Pal** received his M.Sc. (Computer Science) from Allahabad University, UP, India (1996) and obtained his Ph.D. in Mathematics from the Dr. R. M. L. Awadh University, Faizabad (2002). He then joined the Dept. of Computer Application, VBS Purvanchal University, Jaunpur as Lecturer. At present, he is working as Sr. Lecturer of Computer Applications. Saurabh Pal has authored a commendable number of research papers in international/national Conference/journals and also guides research scholars in Computer Science/Applications. His research interests include Image Processing, Data Mining and Grid Computing.